\documentclass[10pt]{artikel3}
\usepackage{amsfonts,amsmath,amssymb}

\usepackage[dvips]{feynmp}

\usepackage{epsfig}

\usepackage{float,braket}
\usepackage{times}
\usepackage{subfigure}

\usepackage{graphicx,wrapfig,color}

\usepackage[small,bf]{caption}

\definecolor{Rood}{rgb}{1, 0, 0} 

\begin{document}
\title{\noindent {\bf Physical spectrum from confined excitations in a Yang-Mills-inspired toy model}}
\author{M.A.L.~Capri$^{a}$\thanks{caprimarcio@gmail.com},\; D.~Dudal$^{b}$\thanks{david.dudal@ugent.be},\; M.S.~Guimaraes$^{a}$\thanks{msguimaraes@uerj.br},\; L.F.~Palhares$^{a}$\thanks{leticia@if.ufrj.br},\; S.P.~Sorella$^{a}$\thanks{sorella@uerj.br}\\\\
{\small \textnormal{$^{a}$ Departamento de F\'{\i }sica Te\'{o}rica, Instituto de F\'{\i }sica, UERJ - Universidade do Estado do Rio de Janeiro,}}
 \\ \small \textnormal{\phantom{$^{a}$} Rua S\~{a}o Francisco Xavier 524, 20550-013 Maracan\~{a}, Rio de Janeiro, Brasil}\\
 \small \textnormal{$^{b}$ Ghent University, Department of Physics and Astronomy, Krijgslaan 281-S9, 9000 Gent, Belgium}\normalsize}

\date{}
\maketitle
\begin{abstract}
We study a toy model for an interacting scalar field theory in which the fundamental excitations are confined in the sense of having unphysical, positivity-violating propagators, a fact tracing back to a decomposition of these in propagators with complex conjugate mass poles (the so-called $i$-particles). Similar two-point functions show up in certain approaches to gluon or quark propagators in Yang-Mills gauge theories.  We investigate the spectrum of our model and show that suitable composite operators may be constructed having a well-defined K\"all\'{e}n-Lehmann spectral representation, thus allowing for a particle interpretation. These physical excitations would correspond to the ``mesons'' of the model, the latter being bound states of two unphysical $i$-particles. The meson mass is explicitly estimated from the pole emerging in a resummed class of diagrams. The main purpose of this paper is thus to explicitly verify how a real
mass pole can and does emerge out of constituent $i$-particles that have complex masses.

\end{abstract}

\setcounter{page}{1}

\section{Introduction}
The analytic study of the spectrum predicted by confining theories still remains a major theoretical challenge.
In particular, despite the intense efforts dedicated to investigating Yang-Mills theories and QCD
using different nonperturbative tools, a complete  understanding of the mechanism of confinement is not yet available,
let stand alone a clean-cut theoretical derivation of the spectrum of massive hadrons and glueballs from (almost) massless gluons and quarks. Such efforts can be appreciated from e.g.~tools like Dyson-Schwinger  equations \cite{Roberts:1994dr,Alkofer:2000wg,Bashir:2012fs} or sum rules approaches \cite{Shifman:1978bx,Shifman:1978by,Narison:2002pw}.

During the last decade there have been many interesting developments in the direction of effectively describing
the dynamics of confined degrees of freedom in Yang-Mills theories. Both analytic and lattice methods
were broadly used to investigate the two-point functions of the elementary fields of quarks and gluons, where the Landau gauge-fixing has been extensively employed.
Recently, very precise lattice studies brought the community to a consensus. The gluon propagator
differs sensibly from its perturbative form of the Faddeev-Popov type: in the deep infrared regime
it is characterized by a clear positivity violation \cite{Bowman:2007du,Cucchieri:2003di} and achieves a non-vanishing value at zero momentum in $D=4,3$ dimensions  \cite{Cucchieri:2007md,Bogolubsky:2007ud,Sternbeck:2007ug,Cucchieri:2007rg,Bogolubsky:2009dc,Oliveira:2008uf,Dudal:2010tf,Oliveira:2012eh}, while it vanishes at the origin in momentum space in $D=2$ \cite{Cucchieri:2007rg,Maas:2008ri}.

From a theoretical point of view, the infrared behavior of the gluon propagator has been object of a quite intense debate over many years, requiring a common effort from several groups. Different nonperturbative techniques have been employed, such as: Dyson-Schwinger equations \cite{Cornwall:1981zr,Alkofer:2000wg,Aguilar:2004sw,Aguilar:2008xm, Binosi:2009qm,Boucaud:2008ji,Boucaud:2011ug}, the renormalization group equations \cite{Fischer:2008uz,Weber:2011nw}, effective field theory framework \cite{Tissier:2010ts,Tissier:2011ey},  the Gribov-Zwanziger approach \cite{Zwanziger:1989mf,Vandersickel:2012tz} and its more recent refined version \cite{Dudal:2007cw,Dudal:2008sp,Dudal:2008rm,Dudal:2008xd,Dudal:2011gd}, and other approaches \cite{Frasca:2007uz,Serreau:2012cg}. These efforts have culminated in the so-called decoupling solution for the gluon propagator, which turns out to be in very good agreement with the lattice data in $D=4,3$ dimensions. Moreover, in $D=2$, these methods coalesce towards a scaling type solution which, unlike the decoupling case, vanishes at the origin in momentum space, being in agreement with the numerical data in $D=2$ \cite{Cucchieri:2012cb,Huber:2012zj}.

One of the analytic expressions \cite{Dudal:2010tf,Cucchieri:2011ig,Dudal:2012zx,Oliveira:2012eh} that is capable of precisely
 describing lattice data in the infrared domain in $D=4,3$ dimensions
and that allows for an analytic  treatment of the calculation of Green's functions is provided by the Refined Gribov-Zwanziger (RGZ) scenario. The Gribov-Zwanziger (GZ) action \cite{Zwanziger:1989mf,Vandersickel:2012tz} is associated with a formulation of the gauge path integral that takes into account the presence of Gribov copies. Its refined version \cite{Dudal:2007cw,Dudal:2008sp,Dudal:2011gd} accounts for the presence of dimension 2 condensates
 which are dynamically generated by the restriction to the Gribov region, necessary to deal with the Gribov issue.
Interestingly, the RGZ analytic expressions for the gluon propagator  are in good accordance with the most recent  lattice simulations and can be naturally decomposed into two quasi-degrees of freedom with complex-conjugated squared masses, which have been called $i$-particles \cite{Baulieu:2009ha}.

Despite these significant advances in the description of propagators of confined elementary degrees of freedom,
the  emergence  of a physical spectrum out of the confined elementary  excitations still remains an open question from the analytic perspective. With the aim of taking advantage of these developments in the description of gluon propagators to shed light into
the spectrum problem, encouraging estimates have been found for the spectrum of the lightest glueball states using analytic tools available in the RGZ scenario \cite{Dudal:2010cd}. Moreover, the inclusion of quark degrees of freedom in the RGZ action seems feasible and interesting results concerning also chiral symmetry and the meson spectrum are currently being investigated. In this toy model paper, we therefore set out to test methods which we can then apply at a later stage to a relevant theory as QCD, where the underlying computations ---in contrast with their toy version--- will be much more complicated and harder to control in an amenable fashion. Clearly, we do not envisage to unravel the deep mystery of confinement here, but rather are we more interested in explaining a first step in how to transform eventually in QCD nonperturbative propagators, vertices etc into approximate information on the spectrum using a framework that is capable of effectively confining the elementary degrees of freedom. To our knowledge, in any continuum approach to the bound state problem, one relies on encoding\footnote{In some cases such input can be obtained self-consistently by solving e.g.~Dyson-Schwinger equations or from minimizing quantum effective potentials to obtain nonvanishing vacuum expectation values.} a variety of hard-to-access nonperturbative physics in effective propagators, couplings or vacuum condensates, where these ingredients are then combined in order to get some information on the QCD spectrum, using Bethe-Salpeter equations, moment techniques, the SVZ sum rule machinery, $\ldots$. In such approaches which each have a number of (dis)advantages, the precise mechanisms for confinement are never needed, for example Wilson or Polyakov loops, center vortices, condensation of magnetic monopoles etc do not enter the picture. In all modesty, we never claimed here or in previous works that we solve the confinement problem.

It should be noticed that  a useful\footnote{In the sense that one can compute quantities with it in a controllable fashion.} analytical form of the propagators is crucial in the construction of a framework that can assess these physical questions that are usually not amenable to analytical tools.

In this paper we investigate the effects of interactions in a toy model based on the idea of confinement as implemented via the RGZ propagator decomposed  into $i$-particles. Our goal is to understand how these complex modes may combine themselves to generate physical propagating modes in  the correlation functions of suitable  composite operators ($=~$bound states). Of course, tackling this problem directly in QCD or even in pure Yang-Mills is an involved task, due to the non-Abelian complex nature of the interactions. It is therefore instructive to approach the problem through a toy model, obtaining a qualitative description of the mechanism of the formation of bound states in theories with unphysical elementary excitations.  We will disclose that an interaction can indeed form a bound state in correlation functions of which a priori it is unclear whether they are physical because of their unphysical constituents. In this sense, the current paper complements the earlier work \cite{Baulieu:2009ha} where it was shown that these propagators can be combined at tree level to give composite operators' correlation functions with physical spectral properties.

Previous descriptions in this line have used the propagators with lattice-fitted parameters as nonperturbative objects that carried information about confinement and estimates of the lightest  glueball masses without including interactions explicitly   have been obtained.
In the current work, we shall combine the nonperturbative propagators with interactions defined in a toy model
so that it is possible to resum a whole class of diagrams contributing to the two-point function of composite operators.
The result for the interacting correlator displays a well-defined K\"all\'{e}n-Lehmann spectral representation.  Moreover,  we show explicitly the appearance
of physical poles, associated with bound states of $i$-particles.
This picture provides therefore a concrete qualitative understanding of the dynamical formation of the physical spectrum of a confining theory, defined in terms unphysical, positivity-violating, elementary fields. It is worthwhile to point out here that propagators with complex mass already found usage in QCD spectrum studies before \cite{Bhagwat:2002tx}.

This paper is organized as follows. In the next section the confining toy model of $i$-particles is presented, including in detail our modeling of the interactions and the definition of the physical composite operators. In section \ref{Resum} we describe the nonperturbative computation of the two-point function of a suitable physical composite operator. Results for the  spectrum of bound states in dimensions $D=2,3$ and $4$ are discussed in section \ref{Results}. The last section is then dedicated to conclusions and final remarks.

\section{The confining scalar field theory}
In this section we describe the model to be studied in the following. We start by defining the free sector of the theory highlighting its confining properties and their formulation in terms of $i$-particles. Then we discuss the introduction of interactions, which should be ultimately responsible for generating a \emph{physical} spectrum of bound states.

\subsection{The free confining theory}

The model we will study has its free sector defined in such a way as to display a confining propagator of the type found in the RGZ framework \cite{Dudal:2007cw,Dudal:2008sp,Dudal:2011gd} :
\begin{align}
D(k) = \frac{k^2+m^2}{k^4 + k^2(M^2+m^2) + M^2m^2+2\theta^4}  =  \frac{R_{+}}{k^2 + M_+^2} + \frac{R_{-}}{k^2 + M_-^2}
\label{scalar-prop}
\end{align}
where $M$ and $m$ are real masses, $\theta$ is a real massive parameter and
\begin{align}
M^2_{\pm} = \frac{(M^2+m^2)}{2} \pm \sqrt{\frac{(M^2-m^2)^2}{4} - 2\theta^4    }
\label{ipart-mass}
\end{align}
where $R_{\pm} = \pm\frac{m^2-M_{\pm}^2}{M_{-}^2-M_{+}^2}$. The right-hand side of eq.(\ref{scalar-prop}) suggests that $D(k)$ encodes the propagation of two quasi-modes with potentially complex conjugated poles,  if $(M^2-m^2)^2<8\theta^4$. This clearly unphysical feature is interpreted as a manifestation of confinement. In gauge theories, such type of gluon propagator shows up as a consequence of the restriction of the path integration to the first Gribov region as a way to deal with the gauge fixing ambiguity of having several solutions to the gauge fixing condition, as is the case in the Landau gauge \cite{Gribov:1977wm,Vandersickel:2012tz}. A dynamically improved version of such Gribov propagator, \cite{Dudal:2007cw,Dudal:2008sp}, does describe the top-notch gluon lattice data very well \cite{Dudal:2010tf,Cucchieri:2011ig,Dudal:2012zx,Oliveira:2012eh}.  The restriction to the Gribov region encompasses the introduction of a set of auxiliary fields, whereby the latter partially mix with the gluon field. It is exactly this mixing that can change the pole structure of the gluon, in particular giving two complex conjugate poles in $k^2$ rather than the single perturbative pole at $k^2=0$. The complex-conjugate poles get an interpretation of short-lived unphysical elementary excitations of the gluon field, see \cite{Stingl:1985hx,Stingl:1994nk} where the same type of propagator was found based on a rational approximation solving scheme to the QCD Dyson-Schwinger equations.

The simplest model leading to this type of Gribov propagator is of the form
\begin{align}
S = \int d^D x \frac 12 \psi \left( -\partial^2 + M^2 + 2\frac{\theta^4}{-\partial^2+m^2}  \right)\psi
\label{scalar-nloc}
\end{align}
from which one easily checks that the scalar field $\psi$ has a free propagator $\langle \psi(k)\psi(-k)\rangle = D(k)$. This model can be rewritten in a local form through the introduction of auxiliary fields
\begin{align}
S = \int d^D x \left( \frac 12 \psi(-\partial^2+M^2)\psi - \bar{\phi}(-\partial^2+m^2)\phi +\theta^2 \psi(\phi+\bar{\phi}) + \bar{\omega} (-\partial^2+m^2)\omega\right)
\label{scalar-loc}
\end{align}
where $(\phi, \bar{\phi})$ is a pair of bosonic complex conjugated fields and $(\omega, \bar{\omega})$ is a pair of anticommuting fields, their role is to cancel the determinant of the $(\phi,\bar\phi)$ integral. In our current toy setting, their dynamics will not be important, but in general these auxiliary ghost fields do play a vital role \cite{Vandersickel:2012tz,Dudal:2008sp,Dudal:2011gd,Sorella:2010it}.

A version of this system was studied in \cite{Baulieu:2009ha} with $M=m=0$, corresponding to the original Gribov propagator. Since this is a quadratic action it can be cast in a complete diagonal form through a change of variables  $(\psi, \phi, \bar{\phi}) \to (V, \lambda, \eta)$
\begin{equation}
S = \int d^D x \; \left( \; \frac{1}{2} \lambda (-\partial^2+M^2_{+}) \lambda + \frac{1}{2} \eta (-\partial^2+M^2_{-}) \eta
- \frac{1}{2} V (-\partial^2+m^2) V + {\bar \omega} (-\partial^2+m^2) \omega \; \right) \;. \label{ipart-act}
\end{equation}
where $\lambda$, $\eta$ and $V$ are real fields, with $V$ being the imaginary part of $\phi$.

In order to simplify the analysis, in what follows we shall set $M=m$. In this case the relevant part of the action (\ref{ipart-act})  reduces to
\begin{equation}
S = \int d^D x \; \left( \; \frac{1}{2} \lambda (-\partial^2+m^2+i\sqrt{2}\theta^2) \lambda + \frac{1}{2} \eta (-\partial^2+m^2-i\sqrt{2}\theta^2) \eta
 \; \right) \;. \label{ipart-act2}
\end{equation}
From this expression one immediately sees that the fields $\lambda$ and $\eta$ correspond to the propagation of unphysical modes with complex masses $m^2 \pm i \sqrt{2} \theta^2$. These are the $i$-particles of the model, namely
\begin{align}
\langle \lambda(k) \lambda(-k) \rangle & = \frac{1}{k^2+m^2 + i\sqrt{2}\theta^2} \; \\
 \langle \eta({k}) \eta({-k}) \rangle & = \frac{1}{k^2+m^2-i\sqrt{2}\theta^2} \;. \label{iprop}
 \end{align}
Expression \eqref{ipart-act2} describes  a theory in which the fundamental excitations are not part of the physical spectrum. More precisely, it is not possible to analytically continue the propagators \eqref{iprop} to Minkowski space-time in order to obtain a well defined particle interpretation for the fundamental fields $\lambda$ and $\eta$. In this sense, we might say that the model displays tree-level confinement.

Note that, even though the $i$-particles action \eqref{ipart-act2} has imaginary mass terms, it is Hermitian if we observe that $\lambda^{\dagger} = \eta$, which follows from $\phi^{\dagger} = \bar{\phi}$. As the actions \eqref{ipart-act} and  \eqref{scalar-nloc} are equivalent and the latter is clearly Hermitian, so should the former be.

\subsection{Interactions}

Many properties of the action (\ref{ipart-act2}) were studied in \cite{Baulieu:2009ha}. In the present work we analyze the system including an interaction between the $i$-particles. Despite of the use of unphysical elementary modes, we shall be able to show  that the interacting system does display a spectrum containing meson-like physical states.

In principle there are many ways of introducing interactions in a model with the tree level propagator in eq.(\ref{scalar-prop}).

To that purpose, we first note that the action (\ref{scalar-nloc}) is invariant under the replacement $\theta^2 \to -\theta^2$. In the $i$-particles formulation (\ref{ipart-act2}) this symmetry can be understood by noting that in a path integral formulation $\lambda$ and $\eta$ are just dummy integration variables. In fact, the partition function is easily seen to be left invariant by $\theta^2 \to -\theta^2$ and $\lambda \leftrightarrow \eta$. Therefore, we shall require that the interactions preserve this feature.

The simplest interaction we can add to a scalar model is a quartic coupling. Thus we propose the following $i$-particles interacting model:
\begin{align}
S = \int d^D x \; &\left( \; \frac{1}{2} \lambda (-\partial^2+m^2+i\sqrt{2}\theta^2) \lambda + \frac{1}{2} \eta (-\partial^2+m^2-i\sqrt{2}\theta^2) \eta + g_1 (\lambda\eta)^2 + g_2 (\lambda^4+\eta^4) \; \right) \;, \label{ipart-inter-act2}
\end{align}
where $g_1$ and $g_2$ are real couplings.

It is interesting to see how these interactions reflect on the original formulation (\ref{scalar-loc}) (with $M=m$). If we write $\phi = \frac{1}{\sqrt{2}}( U + iV)$ with $U$ and $V$ real fields, one realizes that $V$ decouples from $\psi$. In fact, it turns out that the sector of the original scalar model corresponding to (\ref{ipart-inter-act2}) is given by
\begin{align}
S = \int d^D x \left( \frac 12 \psi(-\partial^2+m^2)\psi - \frac 12 U(-\partial^2+m^2)U +\sqrt{2}\theta^2 \psi U + \tilde{g}_1(\psi U)^2 + \tilde{g}_2(\psi^4 +U^4)          \right)
\label{scalar-inter-loc}
\end{align}
where  $\tilde{g}_1 = \frac{g_1}{2} -3g_2$ and $\tilde{g}_2 = \frac{g_1}{4} +\frac{g_2}{2}$.  The relation with the $i$-particles, $\lambda, \eta$, is provided by setting
\begin{align}
\psi &= \frac{1}{\sqrt{2}}( \lambda + \eta)\nonumber\\
U &= \frac{i}{\sqrt{2}}( \lambda - \eta) \;.
\label{scalar-ipart-map}
\end{align}
Another way of defining the same physics is to start from the so-called replica model introduced in  \cite{Sorella:2010it}. In this model we consider two copies of the same theory and couple them through a soft term that has the same effect as the imaginary masses discussed above. Consider for instance two scalar fields $\phi_1$ and $\phi_2$, each described by a theory with quartic coupling. In this case,  the replica model is given by the action
\begin{align}
S = \int d^D x \left( \frac 12 \phi_1(-\partial^2+m^2)\phi_1 + \frac 12 \phi_2(-\partial^2+m^2)\phi_2 +  g(\phi_1^4 +\phi_2^4) + i\sqrt{2}\theta^2 \phi_1\phi_2         \right)
\label{scalar-replica}
\end{align}
The fact that a unique mass $m$ as well as a unique quartic coupling $g$ has been employed in expression \eqref{scalar-replica} follows by demanding that the action displays the mirror symmetry \cite{Sorella:2010it}
\begin{equation}
\phi_1 \rightarrow \phi_2  \;, \qquad \phi_2 \rightarrow \phi_1 \;. \label{mirror}
\end{equation}
The last term in expression \eqref{scalar-replica}, which contains the mass parameter $\theta^2$,  implements a soft coupling between the two replica.
 In fact, this action can be cast into an $i$-particles formulation through the change of variables:
\begin{align}
\phi_1 &= \frac{1}{\sqrt{2}}( \lambda - \eta)\nonumber\\
\phi_2 &= \frac{1}{\sqrt{2}}( \lambda + \eta)
\label{scalar-ipart-replica-map}
\end{align}
giving
\begin{align}
S = \int d^D x \; \left( \; \frac{1}{2} \lambda (-\partial^2+m^2+i\sqrt{2}\theta^2) \lambda + \frac{1}{2} \eta (-\partial^2+m^2-i\sqrt{2}\theta^2) \eta
 + 3g (\lambda\eta)^2 + \frac{g}{2} (\lambda^4 + \eta^4) \; \right)
 \;,
\label{ipart-replica}
\end{align}
which describes the same physics as expression (\ref{ipart-inter-act2}), provided one sets $g_1= 6g_2$  in order to implement the mirror symmetry.

\subsection{Physical operators}
The $i$-particles formulation shows clearly that the excitations corresponding to the elementary fields $(\lambda, \eta)$ are  unphysical. Nevertheless, following the construction outlined in \cite{Baulieu:2009ha,Sorella:2010it}, physical states can be introduced by constructing suitable composite operators out of the fields $(\lambda, \eta)$ which exhibit desirable analyticity properties, as encoded in the K\"all\'en-Lehmann spectral representation.

Such composite operators are obtained by requiring that the fields $(\lambda, \eta)$ enter pairwise, {\it i.e.} the operator contains as many fields of the type $\lambda$ as of the type $\eta$. This will ensure that in the corresponding correlation function only complex conjugate pairs of  $i$-particles will propagate in the Feynman diagrams, a property which provides a good analytic structure. In practice, these operators can be obtained by requiring that their correlation function is left invariant by interchanging  $\lambda$ and $\eta$ \cite{Baulieu:2009ha,Sorella:2010it}.

In the present case, the simplest example of a local composite operator with the required physical properties is ${\cal O}(x) =\lambda(x)\eta(x)$. In fact, at lowest order in the interactions ($g_1=g_2 =0$), it is known that this operator has good analytical properties. In \cite{Baulieu:2009ha} it was shown that the correlation function $\langle{\cal O}(k){\cal O}(-k)\rangle_0$  has a well defined K\"all\'{e}n-Lehmann spectral representation:
\begin{align}
{\cal F}_D(k^2) \equiv \langle{\cal O}(k){\cal O}(-k)\rangle_{0,D} = \int \frac{d^Dp}{(2\pi)^D}\; \frac{1}{(k-p)^2+m^2-i\sqrt{2}\theta^2} \frac{1}{p^2+m^2+i\sqrt{2}\theta^2} = \int_{\tau_0}^{\infty} \frac{\rho_D(\tau)}{\tau+k^2}d\tau
\,,
\label{KL}
\end{align}
where the spectral functions $\rho_D$ have the following expressions, see \cite{Baulieu:2009ha} for  details\footnote{A recent numerical treatment of such correlation functions has been presented in \cite{Windisch:2012zd}.}:
\begin{align}
\rho_{D=2} (\tau) &= \frac{1}{2\pi} \frac{1}{\sqrt{\tau^2- 8\theta^4 - 4m^2\tau}}\,,\\
\rho_{D=3}(\tau) &= \frac{1}{8\pi} \frac{1}{\sqrt{\tau}}\,,\\
\rho_{D=4}(\tau) &= \frac{1}{(4\pi)^2} \sqrt{1-\frac{8\theta^4}{\tau^2}-\frac{4m^2}{\tau}}\,.
\label{spectral-func-D}
\end{align}
The threshold $\tau_0$ is in all cases given by
\begin{equation}\label{thres}
\tau_0= 2\left(m^2 + \sqrt{m^4 + 2\theta^4}\right)\,.
\end{equation}
It is important to observe that all spectral functions $\rho_D$ are positive in the corresponding range of integration. This is a necessary condition in order to interpret \eqref{KL} as the propagation of a physical mode in Minkowski space\footnote{It is useful to remind that, once the K\"all\'en-Lehmann spectral representation for a given correlation function has been obtained, the rotation from Euclidean to Minkowski is well defined   \cite{smatrix1,smatrix2,Itzykson:1980rh}.},  given that upon switching on interactions, the existence of a pole that corresponds to a propagating physical particle still needs to be  established. However, the analyticity properties of the one-loop diagram discussed above will turn out to play a role, hence the interest in this to begin with.

Notice also that, in  $D=4$, the integral \eqref{KL}  is divergent in the ultraviolet, meaning that the correlation function has to be properly renormalized. As it is customary when dealing with spectral representation, this can be implemented be employing a subtracted dispersion relation \cite{smatrix1,smatrix2,Itzykson:1980rh}. In particular, in our case, it will be sufficient to perform only one subtraction and consider the  subtracted expression ${\cal F}_{D=4}(k^2) - {\cal F}_{D=4}(0)$.

\section{Nonperturbative propagator of the composite operator and physical spectrum}\label{Resum}

The next step in establishing a physical spectrum for this confining theory is to compute the correlation functions of the composite operator ${\cal O}(x) =\lambda(x)\eta(x)$ for the interacting case. We shall see that the interactions between $i$-particles will generate a physical bound state, {\it i.e.} a pole  on the negative real axis in the complex $k^2$ plane. This composite correlator featuring a real pole may then be properly rotated to Minkowski space \cite{smatrix1,smatrix2,Itzykson:1980rh}, corresponding to a physical propagating state of the theory.

Since a singularity is not generated at any fixed order of perturbation theory, one must work in a nonperturbative framework. Our approach will be inspired by the standard resummation of bubble diagrams, which is an exact result in $O(N)$-symmetric scalar theories with a large number of field components $N\to \infty$ \cite{Moshe:2003xn}. It should be noticed, however, that the confining toy model discussed in this work is intrinsically different from $O(N)$ scalar theories, since it involves a single doublet of $i$-particles, presenting different (complex-conjugated) masses. We are thus not working in a large $N$ approximation. One may consider our result as that of the analysis of the full theory, but coming from a Bethe-Salpeter approach\footnote{We remind here that the Bethe-Salpeter (BS) equation is the relativistically invariant framework to discuss the binding of 2 particles into a bound state in the context of quantum field theory. In the specific case of QCD, the equation requires the introduction of (in principle) fully nonperturbative input of gluon and/or quark propagators, as well as a sufficiently strong interaction kernel in the equation that is responsible for the binding. Typically, a massive gluon of some sort is assumed \cite{Roberts:1994dr}. Usually, concrete knowledge about this kernel is lacking and one is forced to model/guess. There is no essential difference a priori between solving for the BS amplitude of a bound state and an analysis aiming at computing the correlation function of composite operators with the correct quantum numbers. After all, the main ingredients are the propagators and (highly) nontrivial interaction kernels involved. A simple example: for a quark-antiquark (= meson) bound state in the BS approach, one needs a viable 4-quark interaction mediated by gluons for the BS kernel, but the latter will also be the main ingredient behind the gauge invariant meson correlation function. All physical information of the meson that can be extracted from the BS amplitude can equally well be extracted from the correlator functions, or in particular from their spectral properties as the two ought to be closely related. Let us refer to \cite{Roberts:1994dr,Alkofer:2000wg,Bashir:2012fs}for extensive reviews.} with a bubble approximation for the kernel.

The bubble diagram resummation for the composite correlation function $\langle \lambda(x)\eta(x)\lambda(y)\eta(y)\rangle$ in the confining theory defined in Eq.(\ref{ipart-inter-act2}) is an exact result if one fixes $g_2=0$, ignoring quartic self interactions. Setting $g_2=0$ is in principle a well-defined procedure in dimensions $D=2$ and $3$, since the resulting theories are UV finite and stable under quantum corrections. In the $D=4$ case, (divergent) quartic self interactions are generated via radiative corrections and these terms are in principle needed to guarantee renormalizability. The toy model in $D=4$ with $g_2=0$ that we will solve exactly is therefore either a fine-tuned renormalizable field theory or an effective low-energy theory, protected in the UV by an energy cutoff. In any case, the procedure to be presented in what follows illustrates successfully how an interacting confining theory defined in terms of unphysical degrees of freedom may generate a physical spectrum of bound states in different space time dimensions.

\begin{fmffile}{fmftoy}

Let us now compute the exact two-point function of  the composite operator ${\cal O} =\lambda\eta$. As discussed above, we work with $g_2=0$ so that the only interaction vertex is $g_1 (\lambda\eta)^2$.
The series of diagrams contributing to $\langle {\cal O}(x){\cal O}(y)\rangle$ is:
\begin{eqnarray}
\langle
{\cal O}(x)
{\cal O}(y)
\rangle
&=&\quad
\parbox{2mm}{
\begin{fmfgraph*}(18,18)\fmfkeep{F}
\fmfpen{0.8thick}
\fmfleft{i} \fmfright{o}
\fmf{plain,left,tension=.8}{i,o}
\fmf{dashes,left,tension=.8}{o,i}
\fmfv{decor.shape=circle,decor.filled=empty, decor.size=2thick}{i,o}
\fmflabel{$x$}{i}
\fmflabel{$y$}{o}
\end{fmfgraph*}}
\quad\quad\quad
+\quad\quad
\parbox{7mm}{
\begin{fmfgraph*}(36,36)\fmfkeep{2bub}
\fmfpen{0.8thick}
\fmfleft{i} \fmfright{o}
\fmf{plain,left,tension=.8}{i,v,o}
\fmf{dashes,left,tension=.8}{o,v,i}
\fmfdot{v}
\fmfv{decor.shape=circle,decor.filled=empty, decor.size=2thick}{i,o}
\fmflabel{$x$}{i}
\fmflabel{$y$}{o}
\end{fmfgraph*}}
\quad\quad\quad
+\quad\quad
\parbox{13mm}{
\begin{fmfgraph*}(54,54)\fmfkeep{3bub}
\fmfpen{0.8thick}
\fmfleft{i} \fmfright{o}
\fmf{plain,left,tension=.8}{i,v1,v2,o}
\fmf{dashes,left,tension=.8}{o,v2,v1,i}
\fmfdot{v1,v2}
\fmfv{decor.shape=circle,decor.filled=empty, decor.size=2thick}{i,o}
\fmflabel{$x$}{i}
\fmflabel{$y$}{o}
\end{fmfgraph*}}
\quad\quad\quad
+\cdots
\,,
\end{eqnarray}
where the first diagram corresponds to the free case discussed in the previous section and defined as ${\cal F}_D(k^2)$ in momentum space, cf. eqs. \eqref{KL}. Here, solid/dashed lines represent $\lambda$/$\eta$ propagators, full dots correspond to $g_1$-vertices and the empty dots make explicit the absence of interactions at the extreme points $x,y$, corresponding to the insertions of the operators ${\cal O}(x),{\cal O}(y)$. We omitted all self-energy corrections to the bubble propagators under the working hypothesis that these are already good approximations to the full nonperturbative propagators\footnote{This mimics the observation that the tree level RGZ propagators can account well for the lattice gluon propagator for example \cite{Cucchieri:2011ig,Oliveira:2012eh}. }.

Since the $n$-th term in this series has the general form
\begin{eqnarray}
\overbrace{
\parbox{7mm}{
\begin{fmfgraph*}(36,36)\fmfkeep{nbub-I}
\fmfpen{0.8thick}
\fmfleft{i} \fmfright{o}
\fmf{plain,left,tension=.8}{i,v,o}
\fmf{dashes,left,tension=.8}{o,v,i}
\fmfdot{v,o}
\fmfv{decor.shape=circle,decor.filled=empty, decor.size=2thick}{i}
\fmflabel{$x$}{i}
\end{fmfgraph*}}
\quad\quad\,\cdots
\;\;
\parbox{2mm}{
\begin{fmfgraph*}(18,18)\fmfkeep{nbub-II}
\fmfpen{0.8thick}
\fmfleft{i} \fmfright{o}
\fmf{plain,left,tension=.8}{i,o}
\fmf{dashes,left,tension=.8}{o,i}
\fmfdot{i}
\fmfv{decor.shape=circle,decor.filled=empty, decor.size=2thick}{o}
\fmflabel{$y$}{o}
\end{fmfgraph*}}
\quad
}^{n~{\rm  bubbles}}
\quad
&=&
\int \frac{d^Dk}{(2\pi)^D} {\rm e}^{ik\cdot (x-y)}
\,
{\cal F}_D(k^2)\, \left[g_1{\cal F}_D(k^2) \right]^{n-1}
\,,
\end{eqnarray}
the full two-point function can be written exactly as the result of a geometric series with ratio $g_1 {\cal F}_D(k^2)$. In momentum space, we have
\begin{eqnarray}
\langle  {\cal O}(k){\cal O}(-k)\rangle
&=&
{\cal F}_D(k^2) ~\sum_{n=0}^{\infty}\left[g_1 {\cal F}_D(k^2)\right]^n
=
\frac{{\cal F}_D(k^2)}{1-g_1 {\cal F}_D(k^2)}
\,,
\label{resummed}
\end{eqnarray}
where the difference in the denominator points out to the possible existence of a pole for a given $k^2=-\mathcal{M}^2$.

The explicit conditions for the existence of a pole in the result for the correlator in Eq.(\ref{resummed}) can be found by expanding
around $k^2=-\mathcal{M}^2$:
\begin{eqnarray}
\langle  {\cal O}(k){\cal O}(-k)\rangle
&=&
\frac{{\cal F}_D(-\mathcal{M}^2)+O\left(k^2+\mathcal{M}^2\right)}{
1-g_1 {\cal F}_D(-\mathcal{M}^2)
-g_1 {\cal F}_D'(-\mathcal{M}^2)\, (k^2+\mathcal{M}^2)
+
O\left([k^2+\mathcal{M}^2]^2\right)
}
\,.
\label{resummed-exp}
\end{eqnarray}
Therefore, if the conditions
\begin{eqnarray}
&{\rm (i)}& 1-g_1{\cal F}_D(-\mathcal{M}^2) =0\,,
\label{poleEq}
\\
&{\rm (ii)}& R_D(\mathcal{M}^2)\equiv
-\,\frac{1}{g_1^2{\cal F}_D'(-\mathcal{M}^2)}>0\;,\; {\rm i.e.} \;\;  {\cal F}_D'(-\mathcal{M}^2)<0
\label{Rescond}
\end{eqnarray}
are satisfied, then the two-point correlator of the composite operator $\cal O$ assumes the form of a physical pole at $\mathcal{M}^2$, with a positive residue $R_D(\mathcal{M}^2)$:
\begin{eqnarray}
\langle  {\cal O}(k){\cal O}(-k)\rangle
&\stackrel{k^2\approx-\mathcal{M}^2}{=}&
\frac{R_D(\mathcal{M}^2)}{k^2+\mathcal{M}^2}
\,,
\end{eqnarray}
which corresponds, of course, to a well-defined K\"all\'{e}n-Lehmann spectral representation.

 \end{fmffile}

\section{Results and discussion}\label{Results}

Given the results for the spectral representation of the correlator $\langle{\cal O}(k){\cal O}(-k)\rangle$ in the free case, Eqs.(\ref{KL})--(\ref{spectral-func-D}), one may systematically solve the pole conditions in Eqs. (\ref{poleEq}) and (\ref{Rescond}) in order to obtain the physical spectrum of the confining theory for each set of values for the parameters $(D,g_1,m,\theta)$. All results are shown in this section as dimensionless ratios, with the threshold for two-particle production, $\sqrt{\tau_0}$, being the mass unit used for normalization.

In what follows we analyze three types of massive theories, namely (i) nonconfining ($m\ne 0$, $\theta=0$); (ii) GZ confining ($m= 0$, $\theta\ne0$) and (iii) RGZ confining ($m\ne 0$, $\theta\ne0$). Bound state solutions are found for all cases in all dimensions investigated, $D=2,3$ and $4$.

\begin{figure}[h]
\center
\begin{minipage}{80mm}
\includegraphics[width=8cm]{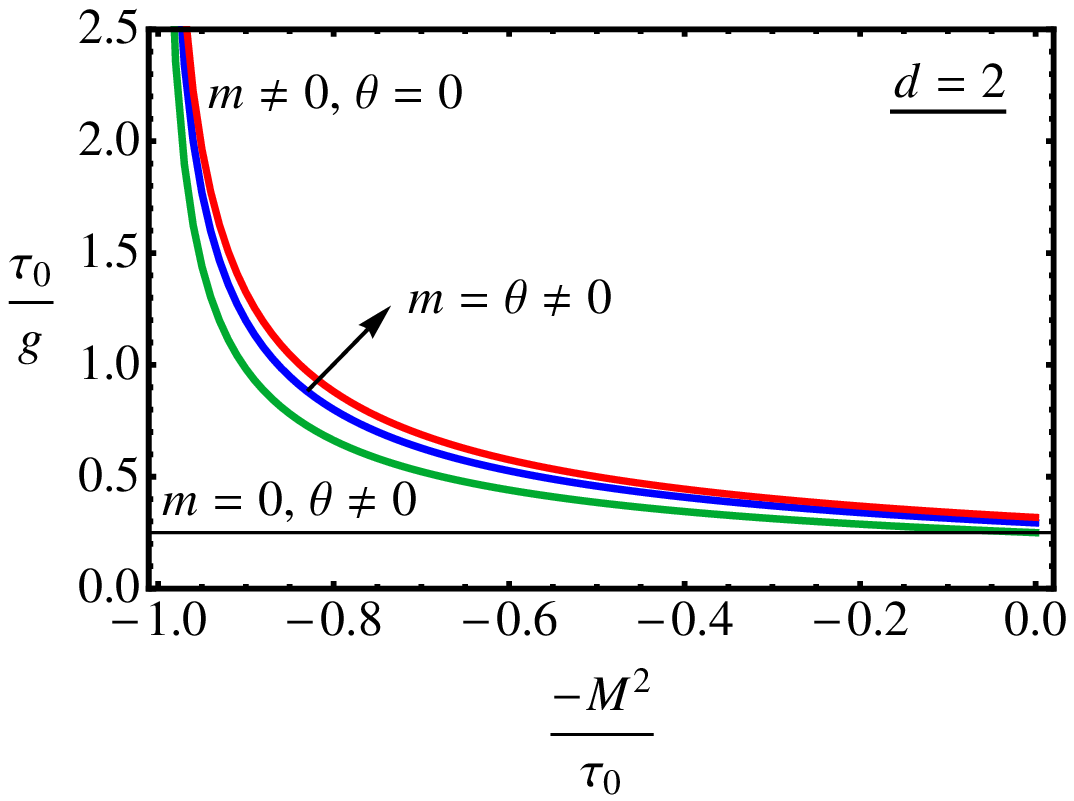}
\caption{Inverse coupling as a function of the (negative) squared mass of the bound state in dimension 2.}
\label{BS-d2}
\end{minipage}
\hspace{0.8cm}
\begin{minipage}{60mm}
\includegraphics[width=6cm]{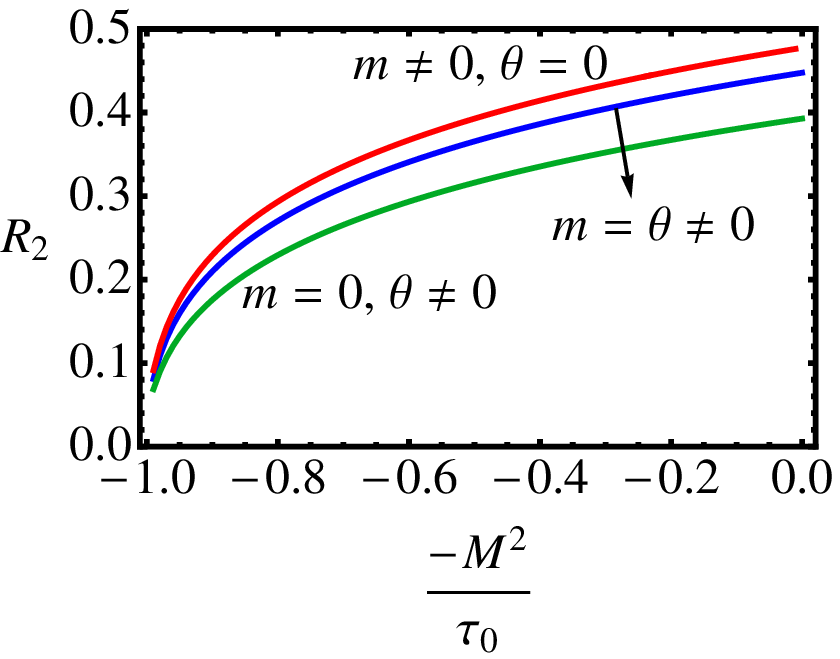}
\caption{Residue of the pole as a function of the (negative) squared mass of the bound state in dimension 2.}
\label{Res-d2}
\end{minipage}
\end{figure}
\begin{figure}[h]
\center
\begin{minipage}{80mm}
\includegraphics[width=8cm]{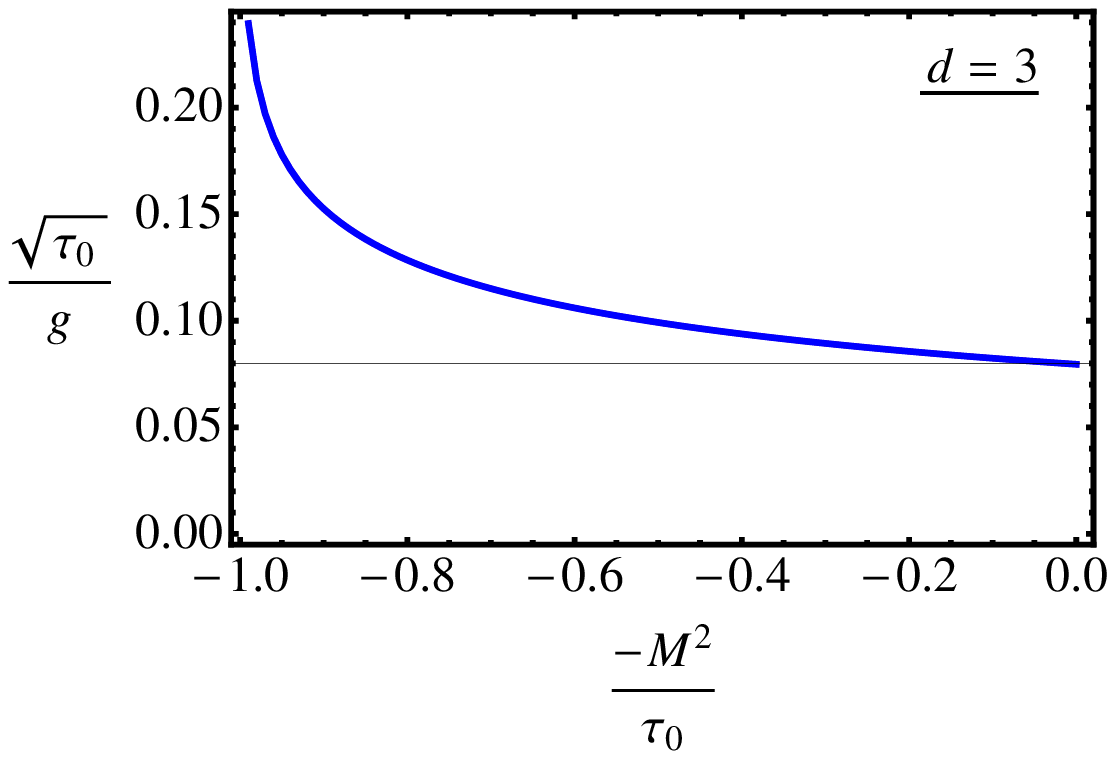}
\caption{Inverse coupling as a function of the (negative) squared mass of the bound state in dimension 3. Here all three theories give the same curve in this normalization scheme.}
\label{BS-d3}
\end{minipage}
\hspace{0.8cm}
\begin{minipage}{60mm}
\includegraphics[width=6cm]{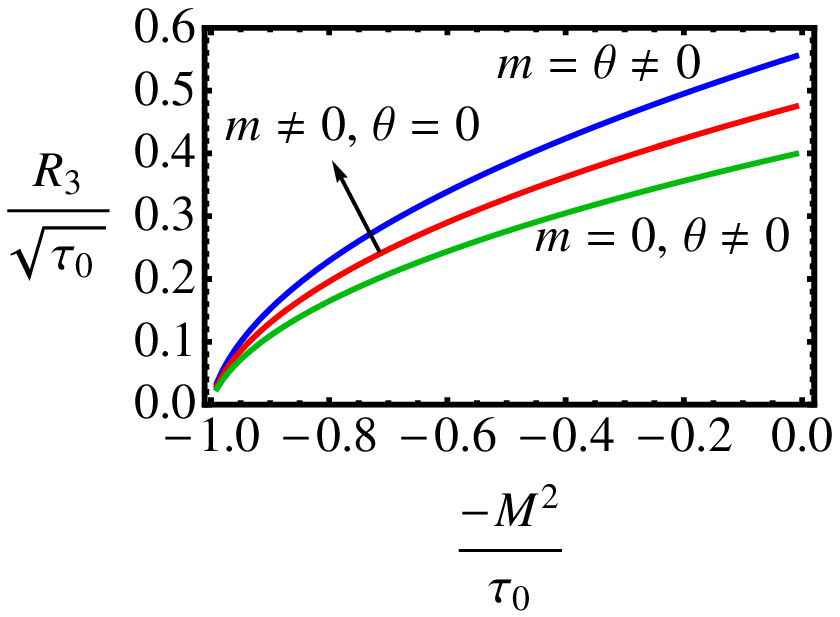}
\caption{Residue of the pole as a function of the (negative) squared mass of the bound state in dimension 3.}
\label{Res-d3}
\end{minipage}
\end{figure}
\begin{figure}[h!]
\center
\begin{minipage}{80mm}
\includegraphics[width=8cm]{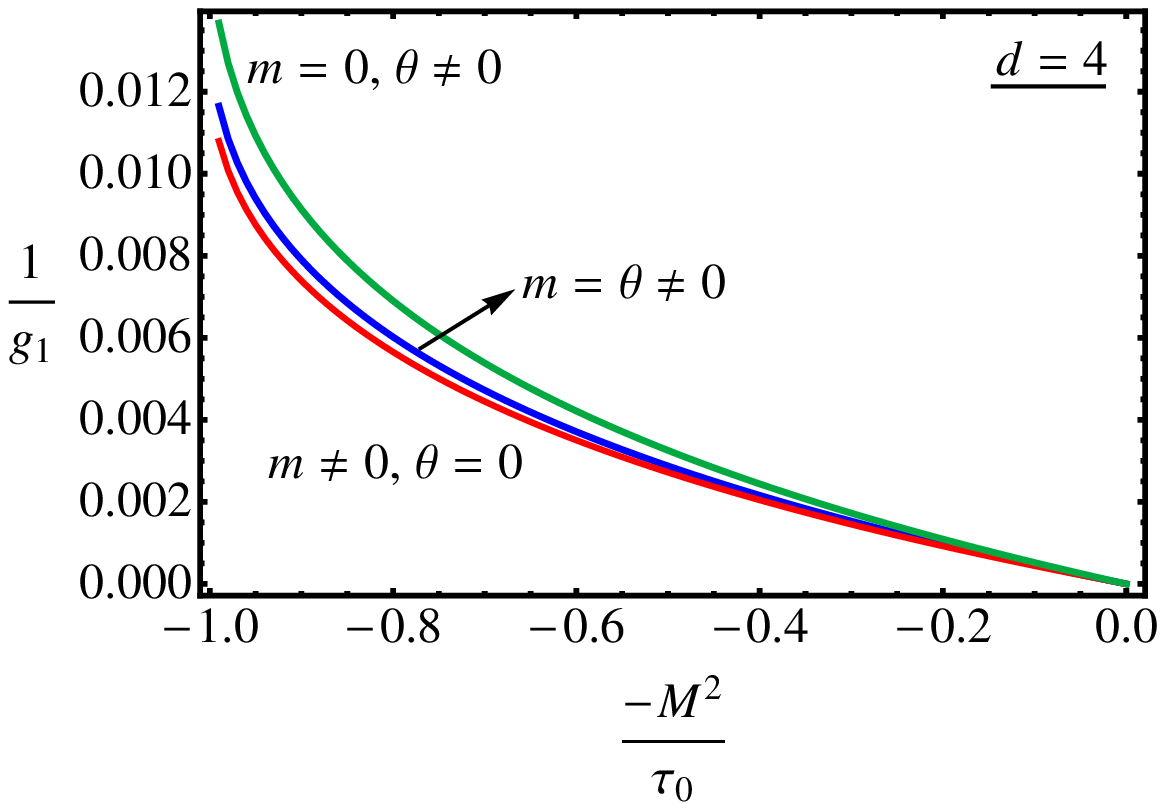}
\caption{Inverse coupling as a function of the (negative) squared mass of the bound state in dimension 4.}
\label{BS-d4}
\end{minipage}
\hspace{0.8cm}
\begin{minipage}{60mm}
\includegraphics[width=6cm]{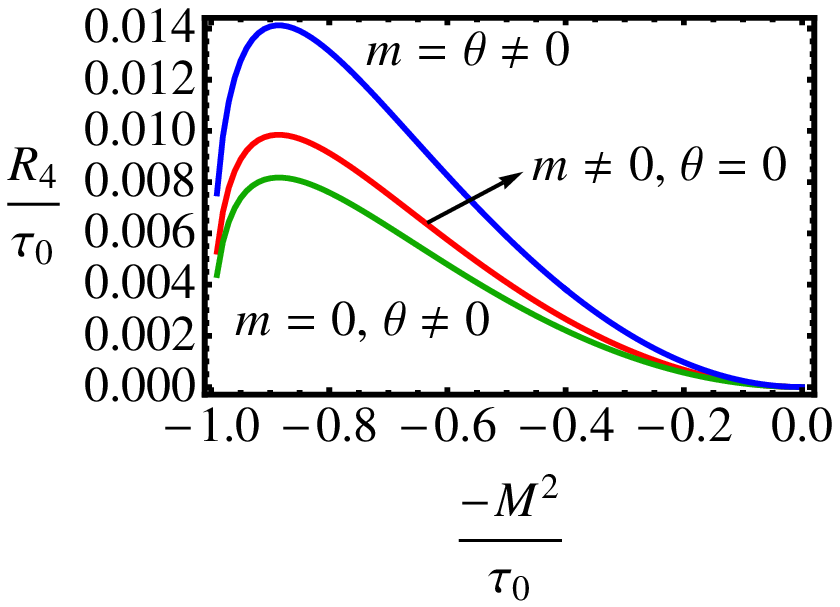}
\caption{Residue of the pole as a function of the (negative) squared mass of the bound state in dimension 4.}
\label{Res-d4}
\end{minipage}
\end{figure}

The mass $\mathcal{M}$ of the bound state is of course a function of the mass parameters $m$ and $\theta$ and the coupling $g_1$, which has positive mass dimension in $D<4$. Its specific form depends moreover on the space time dimension $D$. Nevertheless, the qualitative picture is essentially the same for the three theories analyzed, nonconfining and confining of the GZ- and RGZ-types, as shown in Figures \ref{BS-d2}, \ref{BS-d3} and \ref{BS-d4}. This is an interesting feature of the bound state spectrum in this toy model: its static properties have no particular trace of the type of composites inside the ``mesonic'' state; both confining and nonconfining low-lying excitations furnish qualitatively equivalent spectra, at least statically.

In Figures \ref{BS-d2}, \ref{BS-d3} and \ref{BS-d4}, one further verifies that the  threshold for two-particle production, $\sqrt{\tau_0}$, is a mass upper limit for the bound states found. The heaviest bound states are generated at very low coupling. As interactions are turned on, the binding energy increases and the bound state mass decreases accordingly. It is also shown in Figures \ref{BS-d2}, \ref{BS-d3} and \ref{BS-d4} that the existence of a bound state with mass which is considerably lower than the threshold $\sqrt{\tau_0}$ requires large couplings. For instance, a bound state with half the mass threshold $\sqrt{\tau_0}$ requires couplings of  $g_1/\tau_0^{2-D/2}\approx2, 10$
and $500$ in dimensions $D=2,3$ and $4$, respectively. Furthermore, nearly massless bound states are found in the deeply nonperturbative regime.

In addition, the results show that the theory is physically meaningful only for sufficiently small couplings, $g_1\le g_{\rm crit}$, with $g_{\rm crit}=1/{\cal F}(k^2=0)$ being associated with the massless bound state solution. For couplings larger than this critical value, tachyonic solutions appear, signaling the fact that the toy model becomes ill-defined in this region of its parameter space.  The emergence of the tachyon might signal the instability of the considered vacuum, akin to what happens in the Gross-Neveu model where a condensation of the relevant composite operator takes place accordingly \cite{Gross:1974jv}. Discussion of this would however lead us too far beyond the scope of this paper. This tachyon solution appears in all dimensions investigated, even though it is not made explicit in Figure \ref{BS-d4}. In dimension $D=4$ the critical coupling is directly related to the subtraction scale and assumes therefore a running form, $g_{\rm crit} (\Lambda)$, with $\Lambda$ being the subtraction energy scale.  In Figure \ref{BS-d4}, $\Lambda$ is fixed, for definiteness, so that $g_{\rm crit} (\Lambda)\to\infty$.
 In dimension $D=2$, also the massless bound state, i.e. $g_1=g_{\rm crit}$ should be treated with some care, as is usual for massless particles in dimension $D=2$ \cite{Coleman:1973ci,Abdalla:1991vua,Morchio:1987bf}.

It is also instructive to investigate the behavior of the residue $R_D(\mathcal{M}^2)$ of the bound state poles found, since it is related to the probability amplitude of finding such a state. Figures \ref{Res-d2}, \ref{Res-d3} and \ref{Res-d4} show the results for dimensions $D=2,3$ and $4$, respectively. In line with the expectation that the bound state should disappear in the absence of interactions, the residue go to zero as one approaches $g_1\to0$. For $D=2$ and $3$ the residue grows monotonically as the mass $\mathcal{M}$ of the bound state is decreased, remaining finite also in the massless limit. In contrast, for $D=4$ the residue goes to zero as $\mathcal{M}\to 0$, indicating that the probability of finding this state vanishes. This feature is, however, highly dependent on our subtraction choice for ${\cal F}_{D=4}$ -- which fixes ${\cal F}_{D=4}(k^2=0)=0$ --so that it hardly represents a general physical result,  but it is rather a quantity that deserves a renormalization group/scheme improved analysis.

 \section{Conclusions}
The mechanism that generates the physical spectrum in a confining theory is an outstanding theoretical problem. Inspired by the advances achieved in the last years in the description of propagators of confined elementary particles in Yang-Mills theories and QCD as well as by the encouraging estimates of glueball masses obtained in \cite{Dudal:2010cd}, we adopted the R(GZ) scenario of confinement in which fundamental excitations have unphysical, positivity-violating propagators
described by the combination of modes with complex-conjugated masses, the so-called $i$-particles. As detailed in the introduction, these RGZ propagators
are in good agreement with very precise lattice data for the two-point functions of  gluons in the deep infrared.

In this paper we constructed a scalar quasi-particle toy model of these $i$-particles with the aim of investigating the role of interactions
in the dynamical formation of the physical spectrum. Using  resummation techniques, we have shown that the {\it interacting} two-point correlator of suitable composite operators in this confining theory displays a well-defined K\"all\'{e}n-Lehmann spectral representation with a physical pole appearing dynamically. This picture provides therefore a concrete qualitative understanding of the dynamical formation of the physical spectrum of a confining theory of the RGZ or GZ types, defined in terms unphysical, positivity-violating, elementary fields  with complex mass.

We have analyzed the theory in space time dimensions $D=2,3$ and $4$, showing results for the behavior of the pole mass $\mathcal{M}$ of the physical bound states generated and the corresponding residues as the parameters of the model are varied. Our findings for these static properties of the composite spectrum of nonconfining and confining theories of two types (GZ and RGZ) have proven to be qualitatively similar. It should be noticed that this analysis is done in the Euclidean theory and no statement can be made concerning nonstatic properties of the bound state solutions found. We expect e.g.~that real-time properties, such as widths and decay rates, of the spectrum of confining $i$-particle theories may be strikingly different from that of its nonconfining analog, with real elementary masses.

Bound state formation has been investigated previously in nonconfining scalar theories in different space time dimensions. The theories investigated in the literature are in general not the one we have analyzed here. In dimension $D=2$, the absence of bound states was proven for weakly coupled $\lambda \phi^4$ theories \cite{Feldman:1974hk}. Our results do not contradict these findings because we are dealing with a different model, with two fields $\lambda$ and $\eta$ that interact as $(\lambda\eta)^2$, so that quartic self interactions are strictly absent. The consideration of the toy model with a nonzero quartic coupling $g_2 (\lambda^4+\eta^4)$ would involve extra classes of diagrams that could possibly bring the residue $R_3$ (cf. Figure \ref{Res-d2}) faster to zero in the weakly coupled domain (in the vicinity of $\mathcal{M}^2=\tau_0$), featuring then the disappearance of bound states as seen in \cite{Feldman:1974hk}. In dimension $D=3$, the existence of a nonzero mass gap for $\lambda \phi^4$ theories \cite{Feldman:1976im} was also obtained in the weakly coupled regime, which is already consistent with our findings for the toy model. Bound states in the large $N$ limit of $\lambda (\phi_i\phi_i)^2$ theories were studied in e.g.~\cite{Abbott:1975bn}, whose results are consistent with our findings. It should be noticed however that, even though the large $N$ limit also involves the resummation of bubble diagrams, the toy model adopted here and its results are intrinsically different from those due to the inequivalent interaction vertices. Therefore, we believe our findings are consistent with the literature for nonconfining theories, besides contributing nontrivially to the investigation of bound state formation in confining theories.

\section*{Acknowledgments}
M.A.L.~Capri, M.S.~Guimar\~aes and L.F.~Palhares are supported by the {\it Conselho Nacional de Desenvolvimento Cient\'{i}fico e Tecnol\'{o}gico} (CNPq-Brazil), while
D.~Dudal is supported by the Research-Foundation Flanders (FWO Vlaanderen). The work of S.P.~Sorella is supported by FAPERJ under the program {\it Cientista
do Nosso Estado}, E-26/101.578/2010.

\end{document}